\documentclass{article}[12pt]
\usepackage{epsfig}     
\usepackage{amsmath}    
\usepackage{amssymb}   
\newcommand{\beeq}{\begin{eqnarray}}     
\newcommand{\eeeq}{\end{eqnarray}}     
\newcommand{\be}{\begin{equation}}     
\newcommand{\ee}{\end{equation}}

\newcommand{\hepth}[1]{{\tt hep-th/#1}}

\newcommand\lmp[3]{{\it Lett. Math. Phys. }{\bf #1} (#2) #3}  

\newcommand\phy[3]{{\it Physica }{\bf #1} (#2) #3}   
\newcommand\npb[3]{{\it Nucl. Phys. }{\bf B #1} (#2) #3}     
\newcommand\npa[3]{{\it Nucl. Phys. }{\bf A #1} (#2) #3}

\newcommand\plb[3]{{\it Phys. Lett. }{\bf B #1} (#2) #3}

\newcommand\prd[3]{{\it Phys. Rev. }{\bf D #1} (#2) #3}     
\newcommand\prl[3]{{\it Phys. Rev. Lett. }{\bf  #1} (#2) #3}

\newcommand\mpl[3]{{\it Mod. Phys. Lett. }{\bf #1} (#2) #3}

\begin{document}
\titlepage      
\begin{flushright}     
    
\end{flushright}      

\vspace*{1in}      
\begin{center}      
{\Large \bf An effective Lagrangian for scalar bound states in a dense medium}

\vspace*{0.4in}      
P. Rembiesa      
 \\      
\vspace*{0.5cm}       
{\it  Physics Department, The Citadel, Charleston, SC 29409, USA \footnote{rembiesap@citadel.edu}}\\

\end{center}      
     
\vspace*{1cm}      
\centerline{(\today)}      
      
\vskip1cm      
\begin{abstract}    

Recent analyses suggest that in TeV scales that will be made accessible at the LHC copious amounts of color scalar parton bound states may be produced.    Would this be the case, the scalars would leave long enough to interact and this could lead to new physics.   These interaction could be direct, but also take place with a mediation of the dense parton medium through which they propagate.  Since multiple processes would have to be included, the latter case is too convoluted to be treated with perturbative methods applied to the Standard Model.    We explore a possibility of describing such interaction by a nonlocal Lagrangian which is an infinite polynomial in the field variables, momentum and mass.  We treat all scalars as identical, use a $O(N)$-symmetric Lagrangian, where $N$ is the number of scalars and discuss the problem in $1/N$ expansion.  Nonrenormalizable by all traditional criteria, such model still requires only a limited set of multiplicative renormalizations, provided that the parameters in the Lagrangian are not treated as an infinite set of independent coupling constants, but as  finite expansion coefficients of the Lagrangian in powers of field variables, mass and momenta.  The necessary constraints on the relative magnitudes of these coefficients can be determined  order by order in a double series in the single coupling constant and $1/N$.
\end{abstract} 
\newpage

\section{Introduction}

~~~~The idea of extending the Minimal Standard Model to include higher symmetries and richer particle spectrum is almost as old as MSM itself.  For example the proposition of a four color symmetry in which lepton numbers are considered as the fourth color dates to 1974 \cite{patisalam}.  The major obstacle in increasing the numbers of particles in the model was however related to the fact that this generally leads to unacceptably strong flavor-changing neutral currents.  This flaw turned out to be absent in the case of minimal flavor violation \cite{mfv}.  It effectively reduces the permitted extension of MSM to fields that couple to partons as scalars that carry gauge or color quantum numbers.  The number of representations permitted by the requirement that scalars Yukawa-couple to fundamental fermions is very limited, and for $SU(3)_C\times SU(2)_L$ they include $(8,2), (6,3), (6,1), (\overline{3},3), (\overline{3},2), (3,2), (\overline{3},1), (1,3), (1,2)$, and $(1,1)$ representations.   Of these, even though present in the liteature from the era of technicolor \cite{farsuss}, the color octet is a particularly intriguing candidate, as it has the same weak quantum numbers as the Higgs doublet \cite{manohar}, \cite{moira}.  At LHC energies, the total cross section for production of such such scalars, with masses $\gtrapprox 1.3 TeV$ is large enough for them to be observed in large quantities as heavy resonances resulting from parton merging. \cite{martynov} 

The color scalar bound states are expected to live log enough to ineract.  It was even suggested that they can produce their own (color  singlet) bound states, \cite{kimmehen}, thus   there is no fundamental argument against the conjecture that scalars from all mentioned representations significantly co-participate in partonic interactions.  For example, it has been shown, on phenomenological grounds, that the decays of the color sextet scalars could be responsible for a significant  contribution to the production of top quark pairs even below the 1 TeV region of scalar mass \cite{sextet}.

Among the allowed vertices involving composite scalars, besides the obvious $qqs$ and $ggs$ couplings we also find the $ggss$ vertex.   This implies the possibility of building purely scalar loops, and also the production of cascades of gluon pairs by energetic color scalars propagating in the parton medium.  The  latter could supplement (but not compete with) the known mechanisms of production of gluon cascades by sequential emissions of single gluons by energetic pomerons.   

A separate question is that of the nature of mutual interaction between colored scalars.  The simplest proposition would be to use the contact interaction potential  term in the form of

$\sum_{ab}g_{ab} (\Phi^\star_a\Phi_a)(\Phi^\star_b\Phi_b)$,

\noindent with the summation carried over all scalars (from all relevant, rather than restricted to one chosen representation).  The weakness of such proposition is that it does not involve indirect scalar-scalar multiple interactions that take place with the mediation of the dense parton medium.   Because of the sheer number of possible processes the problem is too complex to be treated perturbatively, however a nonlocal effective interaction might provide an approximate description of the phenomena involved.

The goal of this contribution is to seek a possible form of such effective interaction.  We will employ a $1/N$ expansion (with $N$ representing the total number of composite scalars, as opposed to the dimensionality of the group representation).  Further, as the first approximation we shall ignore individual differences between the scalars (charge, mass, etc), which are secondary for the problem, and consider a nonlocal, $O(N)$-symmetric interaction of identical neutral fields. 

Our goal is to explore infinite-polynomial interactions of scalar fields (that also include higher-order derivative terms), without treating each coefficient of the expansion as an independent interaction coupling that requires a separate renormalization constant.  Instead, we will treat them as finite Taylor expansion coefficients (in the field, mass and momentum variables) of the total interaction Lagrangian.  In no case these coefficients should be regarded as expansion parameters of a predetermined form-factor that assures finiteness of  the amplitudes.   The procedure will be carried order-by-order in $1/N$ and in compliance with  the standard power counting rules in the  coupling constant.

The form of the suggested Lagrangian is typical for  nonrenormalizable  interactions.  However, since we will be working under a crude assumption that all scalars from several possible representations interact identically, we can not assume that this effective interaction could possibly be valid beyond several orders in $1/N$,  where the approximate $O(N)$ symmetry remains valid.   Therefore we will seek an exact form from of the interaction Lagrangian (in terms of its expansion coefficients in all relevant field, mass and momentum variables) from the requirement that, order by order in $1/N$, $p^2$ and $m^2$,  it is renormalizable in the sense that  it requires only a  finite number of standard renormalization constants, similar to that of the  $\Phi^4$ model.    

We will use the dimensional renormalization technique and will demonstrate that it is possible to algebraically determine the necessary relations between the expansion coefficients that lead to mutual cancellations between the harmless (polynomial) poles.  We also offer an argument that our procedure does not lead to harmful logarithmic poles.  The result is almost unique, in the leading order only one of the coefficients of expansion remains free.  In this sense, the condition that  the number of renormalization constant remains finite defines the form of the integration Lagrangian.  Still,  we do not make a claim that the method presented here is much more than a tool for effective models. 

An alert reader may be alarmed that in our procedures the  order of separating pole parts and taking the large-$N$ limit seems to be reversed. Indeed, in the case of cutoff procedures applied to $U(N)$ models, the limit of large N is  taken at a finite value of the  cutoff parameter.  If the infinite limit of the cutoff parameter is taken  first, the procedure is generally invalidated at any finite (although large) order in $N$.  In the case under consideration the order of operations is immaterial. 

The singularities produced by the dimensional method applied to the proposed scalar $O(N)$ interaction are simple poles with the residues coming from a limited number of diagrams which is much smaller than $N$.   All but one of the summations involved are finite.  The only exception is the well discussed and accommodated 'bubble chain'   of   graphs, but this contribution is of the order of $O(N^0)$.  In higher orders in $1/N$ the procedure is therefore perturbative.  Indeed, the  numbers of contributing graphs increase because of the contributions from diagrams topologically identical to the lower-order ones but with redirected $N$-number flow.  However, this effect is algebraic and is not affected by the change of the order of summation and renormalization.  Further, as just explained, the proposed model itself is not expected to retain its validity in such orders of the expansion.

The question of reconciling the order of resummation and renormalization in the $O(N) \Phi^4$ model within the cutoff, rather than dimensional, method  has already been more formally discussed  leading to similar conclusions.\cite {jakovac}

The choice of the $1/N$ expansion is not accidental; our task would not be possible within confines of the conventional perturbation theory which in every order of the coupling contant  mixes vertices with arbitrary numbers of external legs and any possible  momentum dependence.  In contradistinction, the $1/N$ power counting rules establish unique relations between the order of approximation and the maximum number of external legs in  diagrams.   The infinite summations of graphs in the lowest orders of the expansion provide a nonperturbative ground state which serves as an alternative for the conventional free-field solution.  Last but not least, the powers of $1/N$ provide us with a  finite, unaffected by renormalization procedures, non-running expansion parameter  for order-by-order adjustment of the required magnitudes of the expansion coefficients in the Lagrangian.

We end this introduction with a few remarks on nonrenormalizable interactions which are believed to be able to provide limited but useful insight into the nonperturbative realm outside the limits of the asymptotic freedom.

Already  a few years after the discovery of renormalizability of non-abelian theories, Weinberg argued \cite{weinberg1} that, despite the known problems with infinite number of renormalization conditions, with judicious approach, some nonrenormalizable models may still hold predictive powers if the nonrenormalizability is understood as  'a  necessary evil', in the sense that the solutions to a number of initial terms of the expansion of the effective model mimic the behavior of underlying renormalizable and consistent theories. More recently, the argument was made \cite{weinberg2} that it might be possible 'not to restore renormalizability in the Dyson (dimensional power counting) sense, but to live without it' . With the imposition of proper {\it a priori} constraints on the bare action, a gauge theory that is  nonrenormalizable in the usual sense,  can nevertheless be renormalized by introduction of an infinite number of renormalization terms in the quantum action while maintaining its predictive powers.

In a different context of nonrenormalizability of scalar interactions in more than four dimensions, Klauder  \cite{klauder1} (see also his previous work in Ref. \cite{klauder2}), argued that nonrenormalizable interactions are in fact discontinuous perturbations, and therefore adding counterterms to the original free theory may be an  irrelevant procedure.  Instead, he proposed evaluating solutions about an alternative pseudo-free theory.

\section{The model}

~~~~The Lagrangian density of our model is a generic $O(N)$-symmetric  infinite-polynomial interaction of $N$ real (easily generalizable to complex) scalar fields $ \Phi^a$, 

\be
{\cal L}_0 = \frac{1}{2}\partial^\nu\Phi^a\partial_\nu\Phi^a - \frac{1}{2}m_0^2\Phi^a\Phi^a - {\cal L}_S - {\cal L}_I,
\label{eq:lag}
\ee
\noindent
where ${\cal L}_S$ (for future convenience written as an interaction term) represents the nonlocal part of the propagator,

 \be
{\cal L}_S = \frac{1}{2}  \sum_{k+l=2}^\infty  \frac{1}{k!} \frac{1}{l!} d^{(1)}_{kl}\Phi^a (\overleftarrow{\partial^\nu} \overrightarrow{\partial_\nu})^k (m_0^2)^l (L_0^2)^{k+l-1} \Phi^a \\.
\label{eq:prop}
\ee

 In addition to $\mu$, which is the conventional mass unit scale, $L_0$ in (\ref{eq:prop}) represents the  scale of nonlocality range.   The interaction Lagrangian has the form

\be
{\cal L}_I =Ng_0 \sum_{j=2}^{\infty}\frac{1}{j!}\bigg(  \frac{1}{2N\mu^2}  \sum_{k,l=0} ^\infty \frac{1}{k!} \frac{1}{l!} \Phi^a  D^{(j)}_{kl} \Phi^a \bigg)^j,
\label{eq:int}
\ee

\noindent where

\be
D^{(j)}_{kl}=d^{(j)}_{kl} (\overleftarrow\partial^\nu \overrightarrow\partial_\nu)^k (m_0^2)^l (L_0^2)^{k+l}
\label{eq:ker}. 
\ee

Since (\ref{eq:int}) exhibits all marks of a nonrenormalizable theory, some clarifications are necessary before we continue.  The renormalizable  $\frac{1}{N}(\Phi^a\Phi^a)^2_4$ model has three types of divergent operators and three renormalization constants, $ Z_g$, $Z_m$, and $Z_\Phi$.  In the established notation,

\be
g_0 = \mu^{4-n} Z_gg_R=\mu^{4-n} \bigg[g_R + \sum_{\nu=1, j=\nu}^\infty a_{\nu j} g_R^j \frac{1}{(n-4)^\nu} \bigg],
\label{eq:Zg} 
\ee

\be 
m_0^2 = Z_mm_R^2  =m_R^2\bigg[ 1 +  \sum_{\nu=1, j=\nu}^\infty b_{\nu j} g_R^j \frac{1}{(n-4)^\nu}\bigg],
\label{eq:Zm}
\ee
\noindent
and

\be
Z_\Phi = 1 +  \sum_{\nu=1, j=\nu}^\infty c_{\nu j} g_R^j \frac{1}{(n-4)^\nu}.
\label{eq:ZPhi}
\ee

In the model under consideration, the parameters $a_{\nu j}$, $b_{\nu j}$, and $c_{\nu j}$ are also implicit series in  $1/N$.  For sake of transparency of the notation we will not introduce indices of the expansion in powers of this parameter. 
 
The role of derivative terms in the Lagrangian is not to {\it a priori} introduce functions of momenta that modify the vertex or propagator functions so that the standard power counting rules no longer apply and  the divergence problem softens or disappears,  as in the early 'finite theory without renormalization' concept of nonlocal interactions.  Instead we will start from the working assumption that in the initial order in $1/N$ the asymptotic behavior of the propagators is canonical, and that the local-like power counting rules for the vertices apply in the asymptotic domain.   Then we will show by construction that it is possible to to introduce constraints on the expansion parameters so that thanks to mutual cancellations there is no need for the  proliferation of the number of renormalization constants.   In a sense, the method resembles quantum mechanical procedures in which the physical solutions are found by choosing the expansion parameters of a wave function on the requirement of its normalizability.  We do not make any claims about  consistency of the resummed  "full theory".  While the reality may turn out to be more generous,  we seek only a tool for an effective model that behaves as a renormalizable theory through a finite number of orders of the $1/N$ expansion.  

The computational techniques that we will use are not new; the $1/N$-expansion originates from the seventies, \cite{coleman}, \cite{aks}, the infinite-polynomial Lagrangians (without the derivative terms) were first extensively explored by Schnitzer, \cite{schnitzer} and later  also by this author, who applied the method to the non-leading order and bounded (in the field variable) interactions \cite{rembiesa1}, \cite{rembiesa2}.  The regularization of choice will be dimensional, \cite{diagrammar} even though we will not attempt to provide a proof  that all conditions of consistency are met to all orders.  However we will offer an argument that, due to infinite summations, the procedure is not affected by the problem of logarithmic poles due to the the graph topology which includes the $O((1/N)^0)$ 'bubble-chain' contributions in every diagram category.

\section{The leading order}

~~~~The principal $1/N$ power-counting rule for polynomial $O(N)$ scalar Lagrangians  is that every vertex with $n$ pairs of external legs contributes a factor of $(1/N)^{n-1}$, and every closed loop that involves the summation over the field indices contributes a factor of $N$.\cite{schnitzer}   One consequence of these rules is that no graph with $2n$ external legs can appear in the $(1/N)^{n-1}$ order. 

This property of the $1/N$ expansion is  key for our purpose since, unlike in the conventional perturbation expansion, each subsequent order adds only one new class of divergent amplitudes (with two more external legs than the maximum allowed in the preceding order).  Accordingly, the dominating order contains only the vacuum graphs which we shall ignore here, even though they could be important in cosmological applications.  The leading order (LO) contains only the self-energy functions, the next-to leading order (NLO) additional propagator graphs plus four-point functions, and so forth.

\begin{figure}[h]
\begin{center}
\epsfig{file=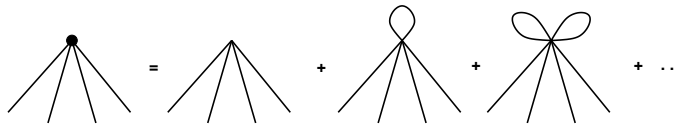,width=0.6\textwidth}
\caption{"Petal" graphs that contribute to the effective four-point vertex.  All vertices in the following graphs are effective vertices, and the heavy dot is not used.}
\label{petal_fig}
\end{center}
\end{figure}

Still, vertices of order higher than $(\Phi^2)^2$ do contribute to the LO, but only as parts of the petal graph series presented in Fig. 1 which form the 'effective' $(\Phi^2)^2$ vertex.  The petal loops do not carry external momenta, and they do not produce overlapping divergences.  Therefore, despite the presence of derivative interaction terms, we will be able to sum them directly.  In purely polynomial models without derivative couplings, such summation is equivalent to taking a derivative of the interaction potential with respect to $\Phi^2$ at a certain value of the effective field.  In the case  under consideration we must proceed in a more deliberate manner.

\begin{figure}[h]
\begin{center}
\epsfig{file=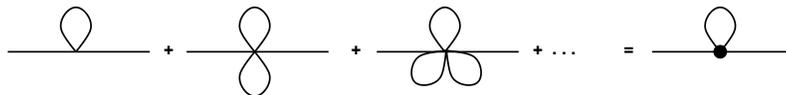,width=0.7\textwidth}
\caption{Daisy graph contributions to the propagator. }
\label{daisy_fig}
\end{center}
\end{figure}

Consider the simplest single-loop propagator diagram shown as the first graph in Fig 2.  It actually represents an infinite sum of graphs  that include vertices with all possible powers of mass and momenta with their respective coefficients $d^{(2)}_{kl}$.  The success of the proposed procedure depends on whether  it is possible to remove all the divergences of these graphs by a single renormalization of mass as in (\ref{eq:Zm}).

Since we are interested only in the general renormalization properties and work on the assumption that the ordinary power counting rules apply in the asymptotics, we will ignore the nonlocal part of the 2-point function and replace it by an ordinary local canonical propagator.  Then the first graph of Fig. 2 is a sum of expressions of the form

\be
I_{k l} =i\mu^{4-n}\frac{g_0}{2 (2\pi)^n}d^{(2)}_{kl}\frac{(L_0^2)^{k+l}}{k! ~l!}\int d^n p\frac{ (p^2)^k (m^2_0)^l}{p^2-m^2_0}.
\label{eq:Ikl}\ee

With help of the identity   $\Gamma(k+\frac{n}{2})\Gamma(1-k-\frac{n}{2})=(-1)^k\Gamma(\frac{n}{2})\Gamma(1-\frac{n}{2})$, and the fact that so far $g_0=g_R$, we find that Eq.(\ref{eq:Ikl}) yields

\be
I_{kl}=\frac{1}{2}\Gamma\bigg(1-\frac{n}{2}   \bigg) \frac{g_R }{(4\pi)^2}m^2_0 \bigg(\frac{m^2_0}{4 \pi \mu^2}\bigg)^{\frac{n}{2}-2} \frac{d_{kl}^{(2)}}{k! ~l!} (m_0^2 L_0^2)^{k+l} .
\label{eq:IEkl}
\ee

Since the last expression involves arbitrary powers of mass with their respective coefficient factors,  there are alternative ways of dealing with the divergence contained in (\ref{eq:IEkl}).  One possibility is to use the fact that $ \Gamma( 1 - \frac{n}{2} )=\frac{2}{n-4} + \gamma -1+O(n-4)$, take

\be 
m^2_0=Z_m  m^2_R=\bigg( 1-\frac{d^{(2)}_{0 0}}{n-4}~\frac{g_R}{(4\pi)^2}\bigg) m^2_R,
\label{eq:Z_m LO}
\ee

\noindent and simultaneously introduce a renormalization constant for the interaction range $L_0$ so that $m^2_0L^2_0 = m^2_RL^2_R$.  

The other option is to retain the condition (\ref {eq:Z_m LO}), but instead of renormalizing $L^2_0$, compensate the effects of higher powers of $m^2_0$ by imposing constraints on the finite coefficients $d^{(n)}_{k l}$.   While the latter approach is more compatible with the proposed technique, it leads to conditions that link $d^{(1)}_{kl}$ to yet unexplored $d^{(2)}_{kl}$.  We  relegate the explanation of  the details of this fact to the Appendix.   

In any case, the necessary constraints on $d^{(n)}_{kl}$  for $n>1$ can be determined solely from the assumption that the asymptotic behavior of the nonlocal propagator is canonical, without possessing any detailed knowledge of its exact form outside the asymptotic region.    The procedure is conditional on whether or not the recurrence  conditions for the coefficients that define the series expansion of the propagator reproduce the canonical behavior at infinity.  Would it not be the case, then  the first procedure would have no alternative.  Nonethless, the canonical behavior of the LO propagator might still be imposed by hand with an appropriate definition of ${\cal L}_S$. 

Hence, we choose to renormalize $L$, and in analogy to (\ref{eq:Zm}) we define 

\be 
L_0^2 = Z_LL_R^2  =L_R^2\bigg[ 1 +  \sum_{\nu=1, j=\nu}^\infty l_{\nu j} g_R^j \frac{1}{(n-4)^\nu}\bigg],
\label{eq:ZL}
\ee

\noindent where in the leading order

\be
l_{11}=- b_{11}. \nonumber
\label{eq:LOZ} 
\ee

\noindent The relation $Z_LZ_m=1$ does not have to hold in higher orders in $1/N$.

In the following we shall follow as closely  as possible the standard renormalization procedures developed for  renormalizable $1/N$ models.  Accordingly, we shall use the $1/N$ expansion for the summation of the dominating classes of graphs while, unless infinite summations of the $O(N^0)$ order are involved,  retain the standard perturbation expansion rules  for discarding  terms of higher  order in $g_R$.   In the case under consideration, the pole parts of all but one $Z_m$ are suppressed by $Z_L$, and the $L_0$  associated with momenta are replaced by $L_R$ since their divergent parts are of higher powers in $g_R$.  Therefore, for the purpose of LO calculations the Lagrangian effectively takes a somewhat simpler form

 \begin{multline}
{\cal L}_S = \frac{1}{2} m^2_0 \sum_{k+l=1}^\infty  \frac{1}{k!} \frac{1}{(l+1)!} d^{(1)}_{k~l+1}\Phi^a (\overleftarrow{\partial^\nu} \overrightarrow{\partial_\nu})^k (m_R^2)^l (L_R^2)^{k+l} \Phi^a \\+ \frac{1}{2}  (\partial^\nu\Phi^a\partial_\nu\Phi_a) \sum_{k=2}^\infty  \frac{1}{k!}  d^{(1)}_{k0}\Phi^a (\overleftarrow{\partial^\nu} \overrightarrow{\partial_\nu})^{k-1}  (L_R^2)^{k-1} \Phi^a.
\label{eq:prop2}
\end{multline}

Also 

\be
D^{(j)}_{kl}=d^{(j)}_{kl} (\overleftarrow\partial^\nu \overrightarrow\partial_\nu)^k (m_R^2)^l (L_R^2)^{k+l}
\label{eq:ker2}.
\ee

Unlike in a conventionally renormalizable $\Phi^4$ interaction,  the mass renormalization alone does not suffice to guarantee the removal of infinities.  The process involves summations of an infinite number of terms that contribute to the residue of the pole, as well as of the finite parts of contributing diagrams.  Further, there are two underlying series summations, one in powers of mass and momenta in the vertex at the base of the loop, and another for a series of daisy diagrams in Figs. 1 and 2, that make up the effective propagator and $(\Phi^a\Phi^a)^2$ vertex.  Both series must be summable.  Otherwise, one would face  infinities of purely algebraic nature which would not be curable by standard field renormalization techniques.  

The total pole part of the sum of all contributions to the first graph in Fig.2 equals

\be
\frac{1}{n-4}m^2_0\frac{g_R}{(4\pi)^2}\sum_{k,l}\frac{d^{(2)}_{kl}}{k!~l!}~(m_R^2L_R^2)^{k+l}.
\label{eq:PP2a}
\ee

 With the simplifying notation

\be
V^{(i)}(p^2,m^2)=\sum_{k,l}\frac{d^{(i)}_{kl}}{k!~l!}(p^2)^k(m_R^2)^l(L_R^2)^{k+l},
\label{eq:DV}
\ee

\noindent  the condition of finiteness of the sum (\ref{eq:PP2a}) translates to the condition that $V^{(2)}(m^2_R, m^2_R)$ is finite.  With the sum of all vertex terms included, instead of (\ref{eq:Z_m LO}), the coefficient of the first term in $Z_m$ would have to be replaced by

\be
b'_{11}=\frac{1}{(4\pi)^2}V^{(2)}(m_R^2,m_R^2).
\label{eq:b11}
\ee

This renormalization guarantees the finiteness of individual petal loops in Fig.2, but the sum of the finite parts of all the daisy graphs 

\be
\Sigma_R=g_R\sum_{j=2}^\infty\frac{V^{(j)}}{(j-1)!}m^2_R\bigg[\frac{m_R^2}{\mu^2}\frac{\gamma-1}{2(4\pi)^2}\sum^\infty_{k,l} \frac{d^{(j)}_{kl}}{k!~l!} (m_R^2 L_R^2)^{k+l}\bigg]^{j-1}
\label{eq:sigma}
\ee

\noindent must also be finite.   This is not a restrictive condition, since it suffices  that  the partial derivative of the interaction potential  with respect to $\Phi^2$ at  $p^2=m^2_R$ is finite.

\begin{figure}[h]
\begin{center}
\epsfig{file=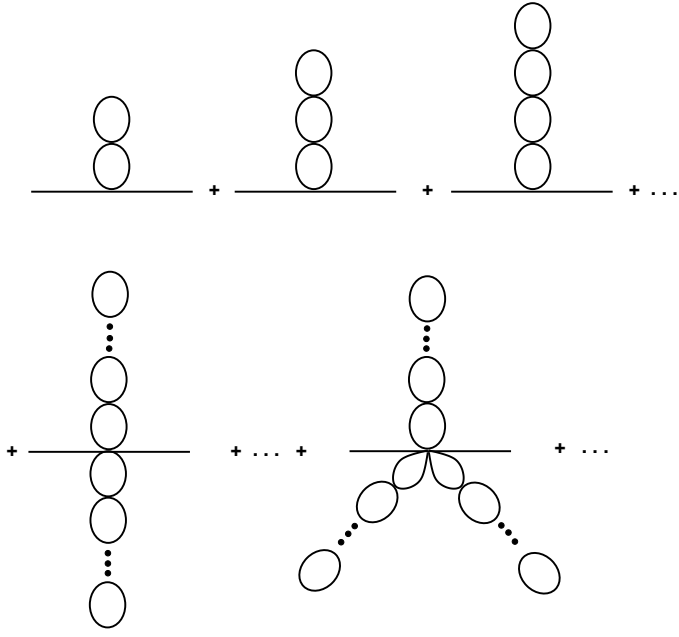,width=0.6\textwidth}
\caption{Stacked-loop contributions to the propagator.}
\label{stacked_loop_fig}
\end{center}
\end{figure}

The LO also includes contributions from the stacked-loop diagrams of Fig.3.  The divergence of the top loop is removed by the just completed mass renormalization, while the remaining loops require renormalization of the four-point function.  Technically, this subject belongs to the NLO, but since NLO graphs are of the order $(1/N)^0$, they can appear as subgraphs in any order.  The contribution to the LO self-energy is simple since the loops do not carry external momenta.

Each loop in the series of stacked diagrams contributes
\be
{\cal G}=\frac{i \mu^{4-n}g_0^2}{2 (2\pi)^n} \sum_{k,l,s,t} \frac{d^{(2)}_{kl}d^{(2)}_{st}}{k!~l!~s!~t!}(m_R^2 L_R^2)^{l+t} \int d^np \frac{(p^2 L_R^2)^{k+s}}{(p^2-m_0^2)^2}.
\ee

 Then, in consequence of the identity 
 $\Gamma\bigg(2-\frac{n}{2}-k\bigg)\Gamma\bigg(k+\frac{n}{2}\bigg)=(-1)^k \frac{\frac{n}{2}-1+k}{\frac{n}{2}-1}\Gamma\bigg(\frac{n}{2}\bigg)\Gamma\bigg(2-\frac{n}{2}\bigg),$  we find that 

\be
{\cal G} = - \frac{g^2_0}{2 (4\pi)^2} \bigg(\frac{m_R^2}{4\pi \mu^2}\bigg)^{\frac{n}{2}-2} \Gamma \bigg(2-\frac{n}{2}\bigg) \sum _{k,l,s,t} \frac{d^{(2)}_{kl} d^{(2)}_{st}}{k!~l!~s!~t!~} \frac{\frac{n}{2}-1+k}{\frac{n}{2}-1}(m_R^2L_R^2)^{k+l+s+t}.
\label{eq:LO4}
\ee

The combined residue, as well as the finite part of (\ref{eq:LO4}) converge as a power series provided that the  sum

\be
\sum_{k,l,s,t} \frac{d^{(2)}_{kl} d^{(2)}_{st}}{k!~l!~s!~t!}(k+s+1)(m^2_RL^2_R)^{k+l+s+t}
\ee

\noindent converges.  Again, this not a very restrictive condition on the specific shape of the interaction potential. It suffices that

\be
{\cal F}(x,y) \equiv \sum_{k,l} \frac{d^{(2)}_{kl}}{(k-1)!~l!}x^ky^l \label{eq:condition}
\ee

\noindent represents an expansion of a regular function that is finite at $x=y=m_R^2L^2_R.$   Then the combined pole parts of (\ref{eq:LO4}) can be cancelled solely by the coupling constant renormalization.  It is worth noticing that the above discussion essentially establishes the sufficient condition for the existence of the effective vertex shown in Fig. 1.

 \section {Four-point function in NLO}
 
~~~~The essential technical element of the $1/N$ expansion is the summation of the infinite series of bubble-chain graphs shown in Fig.5, all of which are of the order of $O((\frac{1}{ N})^0)$.    Let us temporarily postpone the formal question of the chain summation  and first consider the renormalization of individual loops that constitute the chain. 

\begin{figure}[h]
\begin{center}
\epsfig{file=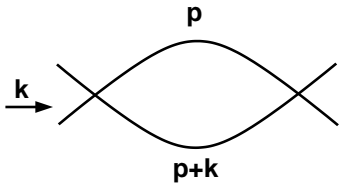,width=0.3\textwidth}
\caption{A single-loop four-point diagram.}
\label{four_point_fig}
\end{center}
\end{figure}

With derivative couplings, there are additional internal momenta factors in loop integrations,  but  the vertices  also provide multiplicative factors of  external momenta that flow in- or out- of the vertex points.  The vertices also contain multiplicative mass factors which combine with those produced by integrations over the internal momenta.  In the graphs under consideration, each internal pair of vertex legs contributes to the numerator of the loop integrand and leads to the integrals of the form 

\be
\frac{i \mu^{4-n} g_0^2}{2 (2\pi)^n} \sum_{klst}\frac {d^{(2)}_{kl} d^{(2)}_{st}}{k!~l!~s!~t! }(m_R^2L_R^2)^{l+t} \int d^nq \frac {[L_R^2 ~q(q+p)]^{k+s}}{(q^2-m_0^2) [(q+p)^2 - m_0^2]}.
\label{eq:oneloop}
\ee

The integral in (\ref {eq:oneloop}) can be written as

\be
i\int d^nq \frac {[q(q+p)]^l}{(q^2 -m^2_0)[(q+p)^2-m_0^2]}=i \int_0^1 dx \sum_{j=0}^l {l \choose j} p^{\mu_1}p^{\mu_2} \ldots p^{\mu_j} I^{jl}_2,
\ee

\noindent where

\be
I^{jl}_\alpha = \int d^n q \frac {q_{\mu_1}q_{\mu _2} \ldots q_{\mu_j}(q^2)^{l-j}}{[M^2-2Pq-q^2]^\alpha}.
\ee

In the last formula we have used a simplifying notation $M^2=m_0^2-p^2x$, and $P=px$.  The singular, as well as the finite parts of the higher-order (in powers of momenta) of the above integrals can be evaluated from the lower-order ones with help of the following recurrence rules for $I^{jl}_\alpha$,

\be
I^{(j+1)l}_\alpha=\frac{1}{2(\alpha-1)} ~\frac{\partial I^{jl}_{\alpha-1}}{\partial P^{\mu_{j+1}}} ,
\label{eq:iter1}
\ee

\noindent and

\be
I^{j(l+1)}_\alpha = -I^{jl}_{\alpha-1} - 2P_{\mu_{j+1}}I^{(j+1)l}_\alpha + M^2 I^{jl}_\alpha.
\label{eq:iter2}
\ee

In the zero order in $m^2L^2$ and $p^2L^2$, the contribution is proportional to 

\be
I^{00}_2=-\pi^{n/2} \Gamma\bigg(2-\frac{n}{2} \bigg) \int^1_0dx(m_0^2-p^2x+p^2x^2)^{\frac{n}{2}-2},
\ee

\noindent with a pole part of $\frac{2\pi^2}{n-4}$.  In the absence of any other divergences, this pole is removed by the standard coupling constant renormalization with

\be
a_{12}=-\frac{1}{(4\pi^2)}(d^{(2)}_{00})^2,
\ee

\noindent which we shall assume to work also for the remaining divergent terms, subject to proper restrictions on the values of the coefficients $d^{(2)}_{kl}$.

In the next (first) order in the combined powers of $m^2L^2$ and $p^2L^2$, the four-point function collects three types of contributions.  One comes from graphs with a factor $m_0^2$ in one of the two vertices, another with a factor of (not subjected to the integration) momenta carried by external legs, and lastly, one with an (integrated) single factor of the product of internal momenta within the loop.  Their combined pole parts expressed in terms of the invariants $m_0^2$ and $(p_1 \cdot p_2) $ equal

\be
\frac{1}{n-4} \frac{g_0^2 ~d^{(2)}_{00}}{(4\pi)^2} L_R^2~[~m_0^2~(2d^{(2)}_{01}+4d^{(2)}_{10})  + 4(p_1 \cdot p_2) d^{(2)}_{10}],
\label{eq:mp1}
\ee

\noindent  where  we have replaced $L_0$ by $L_R$ and $g_0$ by $g_R$ which is exact up to $O(g^3)$,

Despite appearances, there is no need to introduce additional counterterms for the two new divergent operators in (\ref{eq:mp1}).  This is because the Lagrangian of interaction (\ref{eq:int}) already contains terms proportional to the invariants $m_R^2$ and $(p_1\cdot p_2)$ multiplied by $g_0=Z_g g_R$ which also includes the $\frac{1}{n-4}$ factor.  If their residues, and also the  residues of their respective partners from (\ref{eq:mp1}) cancel, no further subtractions are necessary.  The required conditions  are

\be
 d^{(2)}_{10}(d^{(2)}_{00})^2 = 4d^{(2)}_{10}d^{(2)}_{00},
\mbox { and } ~d^{(2)}_{01}(d^{(2)}_{00})^2 =(2d^{(2)}_{01}+4d^{(2)}_{10})d^{(2)}_{00},
\ee

\noindent which are satisfied under only a mildly restrictive condition

\be
d^{(2)}_{00}=4, \mbox{ and }~d^{(2)}_{01}=2d^{(2)}_{10}.
\label{eq:con1}
\ee

The  exact value for $d^{(2)}_{00}$ in (\ref{eq:con1}) should come as no surprise, since the Lagrangian (\ref{eq:int}) contains one free parameter too many ($g$ in addition to the complete set of the $d^{(j)}_{kl}$ coefficients).  

In the second order, the term of the $m^4$ type can be produced by graphs that include the $d^{(2)}_{01}$-type terms in both vertices, or one $d^{(2)}_{02}$-type coupled with a $d^{(2)}_{00}$ partner.  The $p^4$ type requires a pair of $d^{(2)}_{10}$-terms in both vertices, or a  $d^{(2)}_{20}$ factor in one vertex paired with  a $d^{(2)}_{00}$ counterpart in the other.  The latter types contribute also to the $m^2p^2$ factor and combine with contributions from graphs containing pairs of  $d^{(2)}_{10}$ and $d^{(2)}_{01}$, or $d^{(2)}_{11}$ and $d^{(2)}_{00}$.  As before, a proper distinction needs to be made between the graphs in which powers of momenta are applied to the internal legs and affect the loop integration, or apply to external lines and provide only an overall multiplicative factor of external momenta.  Then, by matching the pole residues of the $m^4$, $m^2p^2$ and $p^4$ of the sum of all graphs with similar terms from (\ref{eq:int}), one can determine the values of  $d^{(2)}_{jk}$, where $j+k=2$,  expressed as functions of similar coefficients with $j+k<2$.

This procedure can be extended to higher orders in mass and momentum, and the fact that the value of one seed parameter remains free, guarantees that there exists a degree of diversity in the class of possible solutions.  However,  there are still two possible reasons for the iteration to fail.  One is that, at some order, the system of equations will contradict itself, and the other that for some values of $n,k,l$, the series of the coefficients $d^{(n)}_{kl}$  truncates leading to a simple power of momentum behavior.  Both possibilities can be eliminated by induction.  The argument is straightforward and we will present it in a simplified mathematical notation.

\noindent {\bf i.}~~Assuming that the n-th order in combined powers of $m^2$ and $p^2$ is solved, and the values of $d^{(n)}_{k(n-k)}$ are known, in the $n+1$ order, the equations for $d^{(n+1)}_{k(n-k+1)}$ have the general form 
\be
d^{(2)}_{00}~d^{(2)}_{k~(n+1-k)}=\sum_{j_1+k_1+j_2+k_2=n+1}c_{j_1k_1j_2k_2}d^{(2)}_{j_1k_1}d^{(2)}_{j_2k_2}.
\label{eq:hie}
\ee

\noindent and can be arranged in a hierarchy in decreasing order of the power of momentum $k^2$.  

\noindent {\bf ii.}~~In the top element of the hierarchy,  the coefficient $d^{(2)}_{(n+1)~0}$ appears in the contributing diagrams only once, in the first power, coupled with $d^{(2)}_{00}$ in the opposite vertex.  All the remaining contributions of the same combined dimensionality in $m^2$ and $p^2$ include coefficients of the lower order in $s+k=n$, determined in the earlier steps of the recurrence.  Given the solutions for the lower order, the coefficient $d^{(2)}_{(n+1) 0}$ is therefore uniquely defined.

\noindent{\bf iii.}~~Next equation in the hierarchy  determines the magnitude of the parameter $d^{(2)}_{n1}$.   As in the previous case, there is only one unknown; the right side of (\ref{eq:hie}) includes the term proportional to $d^{(2)}_{(n+1) 0} d^{(2)}_{0 0}$, and all the remaining parameters have been already determined in the lower orders of the iteration.  The steps can be repeated down the hierarchy ladder until the process is completed at $d^{(2)}_{0(n+1)}$.

\noindent{\bf iv.}~~~Since $d^{(2)}_{01}=2d^{(2)}_{10}$, the procedure still contains  one free parameter  $d^{(2)}_{10}$ which sequentially affects the magnitude of every iteratively generated higher-order parameter $d^{(2)}_{kl}$.  This degree of freedom guarantees that it is possible to adjust the magnitude $d^{(2)}_{10}$ to prevent the truncation of the  interaction Lagrangian to a finite polynomial.  This is because the number of conditions is  countable, while the unrestricted coefficient $d^{(2)}_{10}$ is real.

It bears emphasis that, given $d^{(2)}_{01}$, all subsequent $d^{(2)}_{jk}$ are uniquely defined and can be numerically generated with help of Eqs. (\ref{eq:oneloop}) through (\ref{eq:iter2}).  At this time we shall put aside the discussion of the complex question of summability of the resulting multiple series in $k^2$ and $m^2$.

 \section {The bubble-chain revisited}

~~~~Since the graph in Fig.4 is of the order of $O(N^0)$, it appears in the higher-order diagrams only as one of the loops of the arbitrarily long 'bubble chain' shown in Fig. 5.  Typically, the bubble chain is summed as a  Dyson sum, even though (for $g_R>0$) this  generates a Landau-type singularity in the composed $\Phi^2$ field.   The presence of this singularity in the $\frac{1}{N}(\Phi)^4_4$ model was attributed to the expansion around a ground state that is unstable due to the effect of symmetry restoration  \cite{aks}.  

\begin{figure}[h]
\begin{center}
\epsfig{file=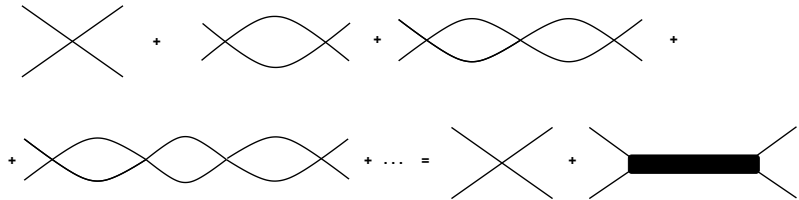,width=0.7\textwidth}
\caption{The bubble chain of graphs. We refer to the rightmost object as to the "incomplete chain".}
\label{bubble_chain_fig}
\end{center}
\end{figure}

With massive fields (and also infinite-polynomial interactions) the pole still appears as a consequence of Borel nonsummability of the underlying series.  The effects of Borel nonsummability were investigated in this context in \cite{rem} and in  \cite{rembiesa2} based on the conjecture that that the Dyson sum is only a representative of a class of possible solutions which differ from each other by a function  whose expansion in powers of the coupling constant is identically zero.  

Ref. \cite{rembiesa2} contains  a  prescription for finding a tachyon-free solution that differs from the (local theory's) Dyson sum by an addition of a nonlocal term which cancels the residue of the pole at $k_0$, but  in the perturbative expansion identically vanishes  term by term.     Since our present objective is restricted to the removal of divergences, it suffices to summarize that it is possible to construct a non-local effective model that is free of the Landau pole and which in the NLO, in the limit of $p\rightarrow \infty$, behaves similarly to the  $\frac{1}{N}(\Phi)^4_4$ model.    Furthermore, since (with the sole exception of the bubble-chain), there is  no further infinite summations in higher orders in $1/N$, the introduction of a correction that eliminates the Landau pole could be regarded as an initial adjustment of the original Lagrangian in order to meet the requirement of consistency.  

An interesting property of the summed bubble-chain is that  it contains a hidden subtraction which can  help with the renormalization procedure.  It can be  best explained on the example of the simpler $\frac{1}{N} (\Phi^a\Phi^a)^2$ model.

If we denote the contribution from a single bubble by $I(k)$ (with all constants absorbed with the only exception of  an overall factor of the coupling $g$), and include in the sum the simple four-leg vertex, the result of the summation is

\be
{\cal G}(k) = \frac {g}{1-g~I(k)},
\label{eq:Dys1}
\ee

\noindent however, when the simple vertex is not included in the summation, the same result can be written as

\be
{\cal G}(k) = g + \frac {g^2I(k)}{1-g~I(k)}.
\label{eq:Dys2}
\ee

Since $I(k)$ grows logarithmically with $k$, in the high momentum regime the right-hand part of (\ref{eq:Dys2}) tends to $-g$. The single difference between the asymptotic behavior of graphs that contain the incomplete bubble chain  and these that include only the simple four-point vertex is in the overall sign factor.  This way the graphs containing the incomplete bubble-chain produce a subtraction for their simple-vertex counterparts.  This kind of subtraction is not enough to reduce the number of divergences, but makes them milder by turning the overall logarithmic divergence of the graph into a double-logarithmic one.   For this reason the standard renormalization procedure still must be implemented.

However,  the hidden subtraction has an important beneficial consequence for the consistency of the  dimensional renormalization which can handle only  harmless poles whose residues are polynomials in masses and momenta, but not their logarithms \cite{diagrammar}. In multi-loop diagrams the elimination of logarithmic poles is possible through mutual cancellations between graphs.   In the early years of the dimensional renormalization, Collins \cite{Collins} verified that in the $\Phi^4$ model the cancellations of logarithmic residues indeed take place on the loop level. 
 
In contradistinction, in $\frac{1}{N}$ models  there is no one-to-one correspondence  between the number of loops and the order in powers of $1/N$. Therefore the loop-level cancellation argument of Ref.  \cite{Collins} no longer applies.   However, the decomposition (\ref{eq:Dys2}) indicates that the logarithmic poles still disappear through the mechanism of implicit internal subtractions.  Every effective four-leg vertex of every graph is made of a simple vertex and a bubble chain which are identical from the viewpoint of the renormalization procedure since their asymptotic forms differ only by an overall sign factor.   Harmless poles can be removed by multiplicative renormalizations which can be evaluated from graphs including only the simple vertices.   The logarithmic poles remain, but they cancel due to the mentioned sign difference.  This effect is distinct, but consistent with the mechanism of mutual cancellations described in Ref. \cite{Collins}.

 \section {Two-point function in NLO}
  
~~~~There are two types of NLO two-point graphs,  shown in Fig.6, that need consideration.  As just explained,all we need to consider are graphs build with only simple four-leg vertices.  When the same renormalization constants are applied to the graphs including the incomplete bubble chain, they cancel their singularities as well. 

\begin{figure}[h]
\begin{center}
\epsfig{file=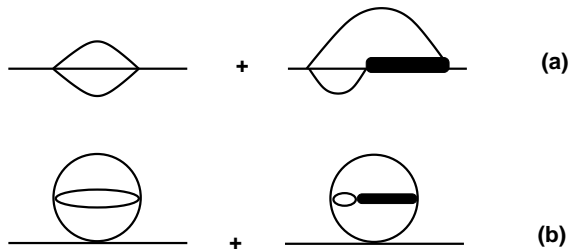,width=0.5\textwidth}
\caption{NLO contribution to the two-point function. The second pair of graphs, shown in (b), is only  an example; additional contributions are generated  by inserting a bubble chain into one of the loops in Fig.~\ref{stacked_loop_fig}.}
\label{NLO_fig}
\end{center}
\end{figure}

Assuming that the asymptotic behavior of the propagator is canonical, in the zero order in combined powers of $m^2$ and $p^2$ in the vertices, the relevant (from the renormalization point of view) part of the two-loop self-energy diagram of Fig. 6a equals

\be
-\mu^{8-2n} g^2_R \frac{(d_{00}^{(2)})^4}{N}  \frac{1}{(2\pi)^{2n}} \int  \frac {d^nqd^np}{(p^2 -m^2_0)[(q+p)^2-m_0^2][(q+k)^2-m^2_0]}, \label{eq:twoloop}
\ee 

The integral over the  Feynman parameters  in  (\ref{eq:twoloop}) contains a pole singularity produced by the divergence of the integrand at equal values of any pair of the parameters involved.  The  harmless poles  of (\ref{eq:twoloop}) are removed by renormalization constants with

\be
c_{12}=\frac{1}{2N}\frac{(d^{(2)}_{00})^4}{(4\pi)^4}, \nonumber
\ee

\be
b_{12}=\frac{\gamma-15}{N}\frac{(d_{00}^{(2)})^4}{(4\pi)^4},
\ee

\be
b_{22}=\frac{6}{N}\frac{(d_{00}^{(2)})^4}{(4\pi)^4}. \nonumber
\ee

\noindent The logarithmic poles cancel by the same mechanism as above.

In the first order in combined powers of mass and momentum one encounters three types of vertex configurations, one with a simple multiplier of $m_R^2L_R^2$, and two with factors including products of momenta.   The extra actors of $p$ originate either from an external leg and an internal propagator, or from a pair of internal propagators.  The numerical value of  the (harmless part of the) pole part of the diagram with a simple mass multiplier equals

\be
 \frac{g_R^2}{(4\pi)^4} L_0^2 \frac{(d_{00}^{(2)})^3  d_{01}^{(2)}}{N}  \bigg[m_0^2 k^2\frac{1}{(n-4)} + m_0^4\bigg( \frac{12}{(n-4)^2} - \frac{16.846}{(n-4)} \bigg) \bigg],  \label{eq:div1}
\ee

The contribution from the diagram in which the momentum factor is produced by the external legs equals

\be
-\mu^{8-2n}2g_R^2L_0^2 \frac{(d_{00}^{(2)})^3d_{10}^{(2)}}{N} \frac{1}{(2\pi)^{2n}}\int\frac{k_\mu p^\mu d^np d^nq}{(p^2-m_0^2)[(q+k)^2-m_0^2][(p+q)^2-m_0^2]}.  \label{eq:external}
\ee

It can be evaluated by an adaptation of the method of Ref. \cite{Collins} which yields 

\begin{multline}
-\mu^{8-2n}\frac{2g_R^2}{(4\pi)^{n}}L_0^2 \frac{(d_{00}^{(2)})^3d_{10}^{(2)} }{N}\Gamma(3-n)k^2 \int_0^1d\alpha d\beta d\gamma\delta(\alpha+\beta+\gamma-1)\frac{ \beta\gamma}{(\alpha\beta+\beta\gamma+\gamma\alpha)^{n/2+1}} \\
\bigg[ \frac{m_0^2 (\alpha+\beta+\gamma)(\alpha\beta+\beta\gamma+\gamma\alpha)-k^2 \alpha\beta\gamma}{\alpha\beta+\beta\gamma+\gamma\alpha} \bigg]^{n-3}.
\label{eq:ppg10a}
\end{multline}

The integrand  in (\ref{eq:ppg10a}) is  singular at $\alpha+\beta=0$  and  $\alpha+\gamma=0$, but the transformation

\be
\alpha=\rho x, ~\beta=\rho (1-x), ~\gamma=1-\rho ,
\ee

\noindent converts the singularity at $\alpha+\beta=0$ into a singularity at $\rho=0$.  In this limit the singular part of (\ref{eq:ppg10a}) equals
\be
(m_0^2)^{n-3} \int_0^1 d\rho dx (1-x) \rho^{1-\frac{n}{2}}=-\frac{(m_0^2)^{n-3}}{n-4},
\ee

\noindent which is the pole part at $\alpha+\beta=0 $.  The pole part at $\alpha+\gamma=0$ is identical, and at $\beta+\gamma=0$   the integral is regular; the pole exists, but at $n=6$.

The divergence  is regularized by subtracting  

\be
m_0^2 \bigg( \frac {1}{\alpha+\beta}+\frac {1}{\alpha+\gamma} \bigg).
\ee

\noindent from the integrand.  The integration can be performed exactly.  The result is somewhat convoluted, but the numerical value of the divergent part of (\ref{eq:external}) equals

\be
\frac{g^2_R}{(4 \pi) ^4} L_0^2  \frac{(d_{00}^{(2)})^3d_{10}^{(2)} }{N} \bigg[k^4 \frac{0.224}{(n-4)}  + k^2m_0^2 \bigg (\frac{2}{(n-4)^2}+ \frac{0.231}{(n-4)} \bigg) \bigg].\label{eq:div2} 
\ee

The last contribution comes from a graph in which the vertex multiplier involves momenta carried by two internal lines over which the field summation is carried out.  The corresponding integral reads

\be
-\mu^{8-2n}\frac{2g_R^2}{(2\pi)^{2n}}L_0^2(d_{00}^{(2)})^3d_{10}^{(2)}\int\frac{p^\mu (p_\mu+q_\mu)d^np d^nq}{(p^2-m_0^2)[(q+k)^2-m_0^2][(p+q)^2-m_0^2]}. \label{eq:internal}
\ee

The integration of the part  that contains $p^\mu q_\mu$ in the numerator can be performed along the same lines.   Its  divergent part  equals

\be
\frac{g^2_R}{(4 \pi) ^4} L_0^2  \frac{(d_{00}^{(2)})^3d_{10}^{(2)} }{N} \bigg[ k^4 \frac{0.448}{(n-4)} + k^2m_0^2 \bigg (\frac{4.000}{(n-4)^2}+ \frac{0.462}{(n-4)} \bigg) \bigg].\label{eq:div3} 
\ee

The case of the part of the integral (\ref{eq:internal}) that contains $p^2$ in the numerator
is different since the integration produces not only $\frac{1}{n-4}$ poles, but also an integral that behaves as  $ \int_0^1d \rho \rho^{1-\frac{n}{2}}$ and is singular in the ordinary sense.   However, the calculation can be easily carried out with use of the decomposition

 \begin{multline}
  \int\frac{p^2 d^np d^nq}{(p^2-m_0^2)[(q+k)^2-m_0^2][(p+q)^2-m_0^2]}= \\
  \int\frac{ d^np d^nq}{[(q+k)^2-m_0^2][(p+q)^2-m_0^2]} + \\
  \int\frac{m_0^2 d^np d^nq}{(p^2-m_0^2)[(q+k)^2-m_0^2][(p+q)^2-m_0^2]},
  \end{multline}   
    
\noindent   The resulting pole part equals
 
\be
\frac{g^2_R}{(4 \pi) ^4} L_0^2  \frac{(d_{00}^{(2)})^3d_{10}^{(2)} }{N}  \bigg[ m_0^4 \bigg( \frac{12.000}{(n-4)^2 }- \frac{11.107}{(n-4)}\bigg) - m_0^2 k^2 \frac{1.000}{(n-4)}\bigg]. \label{eq:div4} 
 \ee

Combining the divergent parts (\ref{eq:div1}), (\ref{eq:div2}), (\ref{eq:div3}), and (\ref{eq:div4}), and using the LO information that  $d_{01}^{(2)}=2d_{10}^{(2)}$ we obtain

 \begin{multline}
\frac{g_R^2}{(4\pi)^4} \frac{(d_{00}^{(2)})^3~d_{10}^{(2)}}{N} L_0^2\bigg[ k^4\bigg( \frac{0.672}{(n-4) }\  \bigg) + k^2m^2_0 \bigg( \frac{6 }{(n-4)^2 } + \frac{1.693} {(n-4)} \bigg)+ \\
m_0^4 \bigg( \frac{36}{(n-4)^2 } + \frac{22.585}{(n-4)} \bigg) \bigg].
\label{eq:divtot}
  \end{multline}
  
  The poles of (\ref{eq:divtot}) need to cancel with analogous terms from the original Lagrangian (corrected with $ Z_g$, $Z_m$, and $Z_\Phi$) and contributions from the LO diagrams  with vertex coefficients $d^{(2)}_{kl}$ so modified that they generate $O(1/N)$ corrections.   So far any restrictions on relations between coefficients $d^{(i)}_{jk}$ were valid in $O((\frac{1}{N})^0)$ order only, and can be adjusted order by order in  higher powers of $\frac{1}{N}$.  In particular, in the NLO

\be
d^{(i)}_{jk}~\rightarrow~\bar{d}^{(i)}_{jk}=d^{(i)}_{jk}+\frac{1}{N}\delta^{(i)}_{jk}, \label{eq:delta}
\ee

\noindent with the sole exception of $d^{(2)}_{00}$ which, as explained in Sec. 4, is exact, hence $\delta^{(2)}_{00}=0$.  In consequence, the LO renormalization constants remain unmodified. 

The residues of  the five poles in (\ref{eq:divtot}) need to cancel with the residues of the poles generated by  graphs represented by the first diagram of Fig. 2, with $\delta^{(2)}_{jk}$ included in in their vertices.  Additionally, conditions relating coefficients $d^{(i)}_{jk}, i\neq 2$ might be imposed as necessary.  Extreme care must be exercised when imposing additional relations between the $d^{(2)}_{jk}$ coefficients, because LO constraints already have been imposed on them in Sec. 4. 

At this point we  need to consider the NLO $O(g_R^2)$ contributions from  $Z_\Phi ~\bar{d}_{20}^{(1)}k^4 L^2_0$,  $Z_\Phi ~\bar{d}_{11}^{(1)}k^2m^2_0 L^2_0$, and $Z_\Phi ~\bar{d}_{20}^{(1)}m_0^4 L^2_0$.  The case of the $k^4$ term is special since  the  LO two-point diagrams do not contribute to the $k^4$ terms.     In consequence, the value of the coefficient $d^{(1)}_{20}$ in the nonlocal propagator must be such that the expression
  
  \be 
 \frac{p^4}{N}(d_{00}^{(2)})^3 \bigg[ d^{(1)}_{20} \bigg(1 +\frac{g^2_R}{(4\pi)^4} \frac{ d_{00}^{(2)}}{2(n-4)} \bigg) +\frac{g^2_R}{(4\pi)^4} \frac{ 0.672~d^{(2)}_{10} }{(n-4)}\bigg]
 \ee
 
  \noindent is finite.  Since  $d^{(2)}_{00}=4$, the pole cancels, provided that
  
  \be
  d^{(1)}_{20}=-~0.336d^{(2)}_{10}.\label{eq:onetwo}
  \ee
  
  While there is no restrictive conditions on $\delta ^{(1)}_{20}$, the coefficients  $\delta ^{(1)}_{11}$, $\delta ^{(1)}_{02}$, $\delta ^{(2)}_{01}$ and $\delta ^{(2)}_{02}$ can be determined from the condition of cancellation of the pole parts and expressed as combinations of the LO coefficients of the Lagrangian.   A straightforward calculation shows that
  
 \begin{align}
  \delta^{(1)}_{11} & = 4.505 d^{(1)}_{11} +4.001 d^{(2)}_{10}=4.505 d^{(1)}_{11} -11.908 d^{(1)}_{20},\nonumber \\ \delta^{(1)}_{20} & =1.634d^{(1)}_{02}-5.609d^{(1)}_{11}-3.383d^{(2)}_{10}=1.634d^{(1)}_{02}-5.609d^{(1)}_{11}+10.068d^{(1)}_{20},\nonumber \\
\delta^{(2)}_{01} & =0.800d^{(1)}_{02} - 1.000 d^{(1)}_{11}+16.375 d^{(2)}_{10}=0.800d^{(1)}_{02} - 1.000 d^{(1)}_{11}-48.735 d^{(1)}_{20}, \nonumber\\ \delta^{(2)}_{10}& =-4.731 d^{(1)}_{11} - 8.009 d^{(2)}_{10}=-4.731 d^{(1)}_{11} + 23.836 d^{(1)}_{20}. \label{eq:deltad}
  \end{align}  
  
 In the limited context of this work there is no need for a detailed discussion of the contributions  generated by the diagram in Fig. 6b.  Its internal lines do not carry external momenta thus the integrations produce a series in powers of mass only.  Mass renormalization can be approached in a manner similar to that used in the case of single-loop contribution from graphs in Fig. 2.   Graphs with all vertex factors of $d^{(2)}_{00}$  define the  $O(g^3)$ contribution to $Z_m$.  Further, in higher orders of $p$ and $m$,  the effect of the insertion of a three-loop 'blob' on a propagator line is the same, independently of the values of $k$ and $l$ in the external $d^{(2)}_{kl}$ factor of the connecting vertex.  In consequence,  just as in the case of the one-loop LO contributions,  it is possible to remove the divergences of all members of this class of self-energy  graphs with a single multiplicative mass renormalization constant.
 
 Another amplitude that will be omitted for sake of brevity is the NLO six-point function which includes three types of graphs.  One is essentially a bubble-chain with one of the four-point vertices replaced by a six-point vertex with extra legs being external.   However, because of the identical topology, the cancellations of the related poles can be produced by a mechanism that closely parallels that which applies to the LO $(\Phi^2)^2$ vertex.   The other graphs are single-loop, with three $(\Phi^2)^2$ or one $(\Phi^2)^2$ and one $(\Phi^2)^4$ vertex that produce six external legs.  The former type of a diagram is finite in the local theory, but here it generates pole singularities due to additional powers of momenta in the vertices.  The residua of all poles can be matched with the expansion coefficients of the $(\Phi^2)^3$ vertex so that the respective singularities cancel.

 \section {Extension to higher orders, conclusions, and speculations}
 
~~~~ Our analysis of the model based on the Lagrangian (\ref{eq:lag}) demonstrates its viability as a candidate for an effective model of interaction of multiple scalars.  The interaction has two components: a contact term (described by the local sector of ${ \cal L}$), and the  nonlocal part of ${ \cal L}$, responsible for interactions mediated by the background parton environment.   While the model was discussed in the context of interactions of colored scalars, other applications are also possible, e. g. scalar bound states in meson sector.  Interestingly, the coupling of such hadronic molecules to mesons can also be modeled by a nonlocal interaction \cite {meson}.  

 We did not attempt to discuss the renormalization properties of the amplitudes beyond the NLO of the $1/N$ expansion, and postponed the investigation of the behavior of the complete series of expansion in powers of momenta.  While the number of free parameters in the model is sufficient to perform the presented procedures, and general patterns revealed in the LO and NLO in $1/N$ leads one to believe that they remain valid also in higher orders, one must be careful not to make overly optimistic claims.   

For example, the assumed property of the bare Lagrangian was that the asymptotic momentum behavior of the propagator is canonical.  For this to be the case, it suffices that in the asymptotic region the nonlocal part of the propagator ${\cal L}_S$, given by Eq. (\ref{eq:prop}) is suppressed by its canonical part, $\frac{1}{p^2}{\cal L}_S(p^2,m^2)\rightarrow0~(p^2\rightarrow0)$, and $\frac{1}{m^2}{\cal L}_S(p^2,m^2)\rightarrow0~(m^2\rightarrow0)$.  There is a danger of inconsistency though, since the requirement of ability to remove all types of divergencies with help of a single $Z_m$ imposes relations between $d^{(1)}_{jk}$ and $d^{(2)}_{jk}$.  Some of the restrictions, like  (\ref{eq:condition}) can be regarded as  only  mild conditions of regularity, but others, like those discussed in Sec. 4 are more restrictive.  While it might be possible to generate a set of $\delta^{(1)}_{jk}$ that would permit cancellation of the poles, maintaing the desired asymptotic behavior of the propagator ${\cal L}_S$ through higher orders in$\frac{1}{N}$is an entirely different question.  

Technically, one could maintain the canonical asymptotics by using unitary transformations to transform the violating terms to the vertices and repeating the procedure order by order in$1/N$.  This however would not resolve any of the formal problems like global unitarity, which would require resummation of the entire series in $p^2$ and $m^2$.  There is a chance that such resummation might (at least in principle) be possible.  The reason is that  for the higher-order vertex to cancel the residues produced by the graphs that include  a product of several lower-order vertices, its single coefficient must be equal but opposite to the sum of product of the lower order coefficients.  The fact that the coefficients are of opposite sign suggests that the signs of the coefficients alternate (since the sums of products involved, this not necessarily has to happen order by order).  Further, the magnitudes of higher coefficients  compare with the products of lower order coefficients.  If these are small, the expansion could possibly be summable within a finite radius of convergence. 

Another essential point is that consistency of the proposed scheme depends on the absence of logarithmic poles.  The argument  for their cancellation  entirely relies on the  presence of bubble chains in diagrams.  Even though in higher orders in $\frac{1}{N}$ it is possible to construct graphs that do not contain a quartic vertex, and hence also no bubble chain exactly as discussed in Sec.5, the procedure still works.  The bubble chain is of the order $O((\frac {1}{N})^0)$ hence, following the prescription illustrated in Fig. 6, it can be attached to every $\Phi^a \Phi^a$ propagator pair in higher-order vertices.

At this juncture we are not entitled to make a claim that we have explained a mechanism of indirect interaction of scalars in a dense medium.  Our goal was to investigate the structure of pole singularities.  The virtue of the presented model is that it permits determination of finite constants in the Lagrangian from the assumption that pole parts of diagrams mutually cancel so that only the standard number of multiplicative renormalization constant is needed to  yield finite results.  The  procedure presented here should be regarded only as a tool for generating an asymptotic series (in $\Phi^a\Phi^a$, $p^2$, and $m^2$) for an effective interaction, particularly applicable in the low-momentum transfer regime.  

Paradoxically, the inclusion of vertices with an unlimited number of external legs, that normally results in nonrenormalizability, was in fact essential for the renormalization procedure.   Since the model was nonlocal, the presence of such vertices should not be interpreted as a dynamical feature of a multiparticle  interaction, but rather as  many-body correlations resulting from a number of intermediate interactions with the medium that eventually affect the scalars.  In a nutshell, correlations between unstable bound states of particles of a renormalizable model should not render it nonrenormalizable.\\\

{\bf Acknowledgement:} This work was supported in part by a grant from The Citadel Foundation.

 \section *{Appendix}
~~~~An alternative method of dealing with the divergences of the LO two-point amplitude is to proceed without  renormalizing the nonlocality range, instead using a procedure applied to the four-point diagram.  We opted for the first alternative for the reason of economy, and because it does not require any upfront constraints on the coefficients of the nonlocal part of the propagator.  This makes it easy tomeet the requirement of the canonical asymptotic behavior of the LO propagator.  The alternative approach would require a prior definition of $d^{(2)}_{kl}$ coefficients which are not determined until the next order of the $1/N$ expansion.  
In the lowest order in $m^2$,  only $I_{kl}$ requires renormalization and the expression (\ref{eq:Z_m LO}) correctly defines $Z_m$.  In the next order one renormalizes the sum of $I_{01} + I_{10}$.  The part of $I_{01} + I_{10}$ that is  relevant  for the renormalization procedure reads

\be
I_{01} + I_{10} =\Gamma\bigg(1-\frac{n}{2}   \bigg) Z^2_m g_R \frac{m^4_R L^2 }{2 (4\pi)^2}  (d_{01}^{(2)}+d_{10}^{(2)}).
\label{eq:2Ekl} 
\ee

 All graphs  involved so far were of the first order in $g_R$,  therefore the factor $Z^2_m$ in (\ref{eq:2Ekl}) can be dropped.  From the conventional point of view, separate renormalization constants for $d^{(2)}_{jk}$ are needed (which would imply  nonrenormalizability in its traditional meaning), or imposition of the condition $d^{(2)}_{01}=-d^{(2)}_{10}$ (which would imply triviality).  However, the propagator (\ref{eq:prop}) also contains the term
 
 \be
 \frac{1}{2} m_0^2 d^{(1)}_{02}\Phi^a  (m_0^2L_0^2) \Phi^a,
\label{eq:term}
\ee

\noindent  and the cancellation of infinities is possible, provided there exists an appropriate relation between $d^{(1)}_{02}$ and $d^{(2)}_{01}+d^{(2)}_{10}$.    In our case this means that $d^{(1)}_{02}d^{(2)}_{00}=d^{(2)}_{01}+d^{(2)}_{10}$.  Order by order, similar relations  between $d^{(1)}_{0n}$ and $d^{(2)}_{j~(n-j)}$ also follow.  However, in order to determine the exact form of the condition responsible for the cancellation of the pole  of mass dimension $(m^2_0)^{s}$, we would need to consider the sum of all expressions (\ref{eq:IEkl}) for which $k+l=s$.  This would lead to a condition linking $d^{(1)}_{0s}$ to a weighted sum of $d^{(2)}_{k~(s-k)}$ which could not uniquely define $d^{(2)}_{kl}$.



\end{document}